**PAPER • OPEN ACCESS**

# Discovering the formal definition of limit through exploration in dynamic geometry environments

To cite this article: Y D Kristanto *et al* 2019 *J. Phys.: Conf. Ser.* **1180** 012004

View the article online for updates and enhancements.

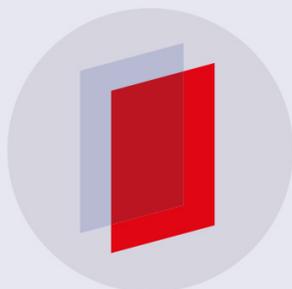

## IOP ebooks™

Bringing you innovative digital publishing with leading voices to create your essential collection of books in STEM research.

Start exploring the collection - download the first chapter of every title for free.

# Discovering the formal definition of limit through exploration in dynamic geometry environments

Y D Kristanto[1], M M Melissa[2], and A H Panuluh[3]

Universitas Sanata Dharma, Paingan, Maguwoharjo, Depok, Sleman, D. I. Yogyakarta, Indonesia

yosepdwikristanto@usd.ac.id

**Abstract**. The formal definition of limit is one mathematical idea that gives difficulties to most students. Therefore, it is necessary to provide engaging activity in order to construct the students' understanding about the definition. This study aims to shed light on designing activities with dynamic geometry software to facilitate the students in discovering the formal definition of limit. Forty-eight students participated in the study. Both quantitative and qualitative data were collected from the students' responses in and attitudes toward the activities. The students' responses were gathered directly by using the dynamic geometry software whereas an online questionnaire was used to collect the students' attitudes. The results of the study then presented in this paper. Further, a set of suggestions in designing and implementing dynamic geometry environments to foster students' understanding about formal definition of limit was also proposed.

## 1. Introduction
Many students consider mathematics as "inhuman" because they have mindset that mathematics far from their daily life. As posited by Freudenthal [1], mathematics educators should instil mathematics to become more humanity. One possible way to accomplish the goal is to make mathematics can be recreated by the students. It is also the case in Calculus learning, especially in the topic of formal definition of limit. On the one hand, the formal definition of limit is one mathematical idea that gives difficulties to most students [2]. On the other hand, the concept is mathematically important. It is arguably the most fundamental concept in Calculus [3]. Students should have understanding on the concept in order to learn other concept in Calculus. Therefore, it is necessary to provide engaging activity in order to construct the students' understanding about the formal definition of definition. One strategy to attain this aim is to provide dynamic geometry environments (DGE).

## 2. Literature review

### 2.1. The definition of limit
Many Calculus textbooks [4–8] present the concept of limit in similar way. First, they introduce limit of functions numerically and graphically, then continue to be more precisely. The use of numerical and graphical approaches is aimed to create concept images of limit. The concept images are important for students in understanding and comprehending the formal definition of limit. Informally, limit of a function *f*(*x*) as *x* approaches *a* equal to *L* if the values of *f*(*x*) close to *L* as *x* sufficiently close to *a* [5,p.51]. This "informal definition" of limit with appropriate design tasks will lead to the concept images of limit creation [9].

In shed light on how limit concept should be understood, Cottrill et al. [10] have proposed a framework in this regard. The framework consists seven steps that can be categorized into two general phases. The first four steps focus on informal limit concept and the remaining three steps emphasis on formal definition of limit. This framework shows the value in bridging the informal and formal concept of limit. The similar perspective also can be found in Zandieh and Rasmussen [11] study. Further, Fernández [12] add that when students were allowed to build on their spontaneous conceptions, they tend to reason more coherently about the formal definition of limit.

Beside the important of task sequencing in understanding the formal definition of limit, it is also important to analyse the concepts in the definition. Formally, limit of $f(x)$ as $x$ approaches $a$ is $L$ if for every number $\varepsilon > 0$ there is a number $\delta > 0$ such that the following statement is true: If $|x - a| < \delta$ then $|f(x) - L| < \varepsilon$. In understanding the definition, students must understand the basic logic, especially for implication. Further, they must also understand the absolute values. These two critical concepts are must mastered by students if they work in problems involving proof. They should proficient on how choose the number $\delta > 0$ such that for every $\varepsilon > 0$, if the distance between $x$ and $a$ is less than $\delta$ then the distance between $f(x)$ and $L$ is also less than $\varepsilon$.

*2.2. Dynamic geometry environment*
Rich body of research has reported the use of DGE in mathematics classroom [13–15]. Chan and Leung [16] conducted systematic review on the effectiveness of DGE on students' achievement. From their studies, it is found that DGE has a positive and large effect size on mathematics achievement. The DGE is suitable with inquiry-based learning, but also others teaching strategies that emphasis teacher as facilitator. Specifically, Jones [17] has reviewed the use of DGE in facilitating students' reasoning in mathematical proof. The exploration and visualisation in DGE might lead the students' understanding of the need for and the roles of proof. Further, the use of DGE for conceptual exploration can result in conceptual gain. However, the DGE itself will does not grant an impact on students' learning. Teacher should facilitate other activities and guidance through the use of DGE in classroom.

Hohenwarter and colleagues [18] gave illustrations on how the dynamic geometry software can be utilized in Calculus classroom. They noted that the DGE can be used in presentation or mathematical experiments. For the presentation purpose, DGE is a part of teacher's lesson planning. Teacher can develop interactive activities or mathematical graphic from their desk, and then present the materials in classroom. In addition, teacher also can construct dynamic figures directly in classroom if he/she want more flexible teaching. The DGE also enable the mathematical experiments to be conducted in classroom. Students can use the dynamic geometry software to construct and explore mathematics graphs in understanding mathematical concept. Further, teacher also can provide premade interactive activities for students to explore and investigate certain mathematical concepts. In these activities, students need to interact in computer environment.

*2.3. Purpose of the present study*
Build from the body of literature about research in limit of functions and DGE, the present study aims to describe the implementation of web-based dynamic geometry environment to facilitate students in understanding the formal definition of limit. This paper also adds the students' perceptions about the DGE for their learning.

**3. Method**
The present study was exploratory research that was conducted at a private university in D. I. Yogyakarta, Indonesia. The study was carried out in Differential Calculus class that contains 51 freshman, sophomore, junior, and senior year students. However, there were 48 students that attended the class when the study was held, namely 13 males and 35 females. All of the students were pre-service mathematics teachers that obliged to pass the Differential Calculus course.

The current study was conducted on one meeting when the class discussed about the formal definition of limit. Before the meeting, the first author as an instructor developed interactive activities with Desmos. The activities contain fifteen slides that can be categorized into three general phases, i.e. introduction, main, and closure sections. The introduction section informed the learning goals and give warming up activities to students. In the warming up activities, students recalled their knowledge about informal definition of limit and basic algebra that were needed in understanding the following body of lesson. In the main section, students practiced to use the formal definition of limit to prove statements about limit concept. In this section, students were guided to choose appropriate a number delta with the assist of dynamic graphics. The dynamic graphics have slider that can be set by students to visualize the implications of their chosen delta. In the closure, students were appreciated after they finished all the activities.

Students carried out the activities in group of three until four. Each group has one laptop that connected to internet network. The students discussed in group to solve the given problems. During the discussion, the instructor visited from one group to another group to guide the groups. The instructor answer questions that arose from the students. The students could see the progress of the other groups' works that were displayed in front of classroom. They also could see the correctness of the works of others and their self, and if they want, they can revise their answers.

## 4. Result and discussion

The first aim of the current study is to describe the implementation of web-based dynamic geometry environment to enhance students' understanding of the formal definition of limit. The proportion of students' participation in each activity's section is presented on Figure 1. The figure reveals that most students have finished the introduction and main activities, whereas it is only one group (5.3%) which has completed the closure activity. In the introduction section, most students did not complete an activity in which they were asked to write the compound inequality into absolute value expression while most students failed to complete an activity in which they should prove a statement using the formal definition of limit in main section.

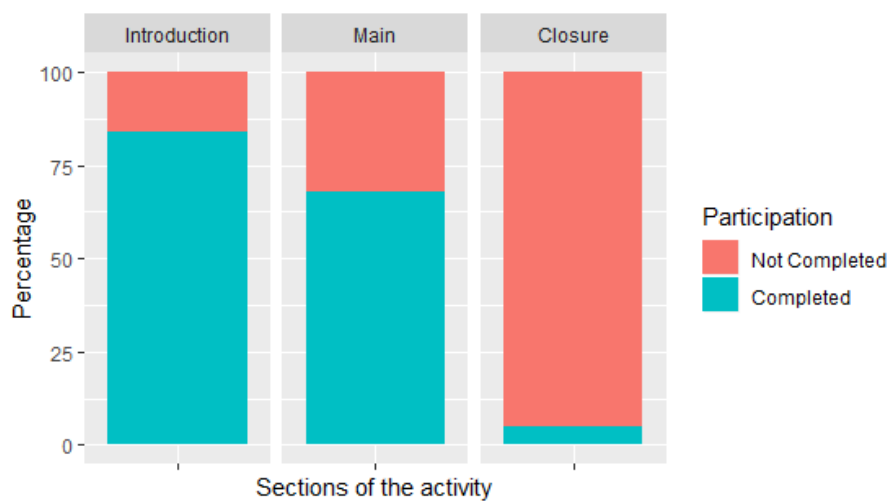

**Figure 1**. Students participation in each section of activity

The aforementioned finding in student participation shows that students still have difficulties in understanding absolute value expression. The absolute value is one of several concepts to be understood in understanding the formal definition of limit. The lack of comprehension in the absolute value concept and others prerequisites will hinder students' success in studying the formal definition of limit [19]. Further, it also found in the current study that many students still lack of skill in using the formal definition of limit to prove a mathematical statement. They tend to use limit rules, e.g. substitution rule, and table to prove the statement instead of using the limit definition.

In addition to students' participation, the present study also gathered the students' perceptions on the DGE. Overall, students have positive attitude on the DGE. They perceived that the DGE help them to learn limit concept (3.73 out of 5), the DGE is easy to use (3.73 out of 5), the DGE is interesting (4.4 out of 5), and they found that the appearance of the DGE is interesting (4.27 out of 5). The students' perception is presented in Figure 2. The similar result also has been found in literature [20–22].

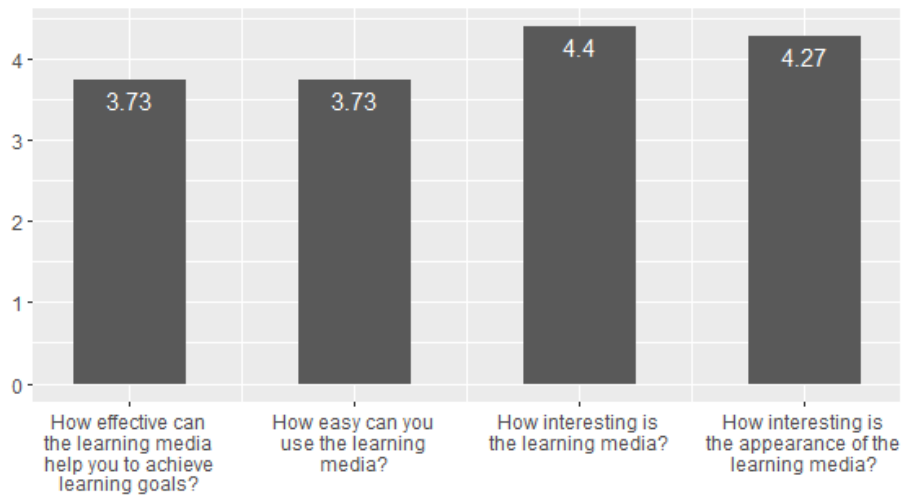

**Figure 2.** Students' perceptions toward the DGE

From the students' answers of open questions in the questionnaire, it can be identified the students' positive attitudes in experiencing the DGE. First, students perceive that the collaborative learning facilitated by the DGE can enhance their learning. They acknowledge that the group discussion helps them to solve the problems given in DGE. One possible explanation of this students' perception has been stated in Roschelle and Teasley's study [23]. They posit that students construct shared conceptual space in collaborative learning. Second, the students notice that the DGE was interesting, easy to use, and has value on its visualization. Third, students appreciate the dynamic feature of the DGE. With slider feature, for example, they can see what happens to epsilon when they change the values of delta. The students' perceptions with regard to the DGE benefits are as follows.

> *[The learning] was interesting and [with DGE] I can provide explanations of the problem's answer to others. ($S_1$)*
>
> *The display is interesting and the graph is easy to understand. It is be caused when I pointing the pointer on the graph, I directly can see what coordinate the point is. ($S_9$)*
>
> *[The positive experience] was when I move the slider, the graphic is also changing. ($S_{26}$)*

Students also had negative experiences in DGE. First, they encountered the technical issues, such as internet connection and devices ownership. Not all students have laptop or tablet to be used to access the DGE. Internet connection is also important when implementing the DGE. The trouble in the internet connection will hinder the students' learning. Second, students found difficulties in typing mathematical equation. The issue arose because the dynamic geometry software (i.e. Desmos) still had limitation in providing mathematical equations. Lastly, students admitted that they have difficulties in understanding the concepts provided in the DGE when they lack of prerequisite knowledge and skills. The importance of prerequisite knowledge and skills is consistent with Dogan-Dunlap [24]. The prerequisites provide learning context to students in knowing and understanding new topic.

## 5. Final remarks

This paper explores how using the DGE to enhance students' learning in Differential Calculus class, especially in the topic of the formal definition of limit. Students' participation in the designed activities were described in the paper. Further, students' positive and negative attitudes toward the activities were identified as well. The findings of the current study give contribution on how implementing DGE to foster students' understanding about formal definition of limit. First, instructor's intervention is needed when implementing the DGE. The DGE itself cannot enhance students' learning. Instead, the instructor is still needed to provide learning environment in which students can learn optimally. Second, the DGE may have positive impact to students' learning when it is implemented in collaborative learning. In this setting, students can discuss each other on the given problems to construct their knowledge and skills. Third, task design in the DGE should be designed appropriately so that can be completed on time. Finally, the instructor should provide pre-instructional activities before implementing the DGE so that students can prepare their prior knowledge and skills.